\documentclass[preprint,showpacs, superscriptaddress, preprintnumbers,amsmath,amssymb]{revtex4}

\usepackage{booktabs}
\usepackage{color,graphicx}
\usepackage{amsmath}
\usepackage{dcolumn}
\usepackage{CJK}                 
\usepackage{bm}                  
\usepackage{soul}
\usepackage{enumerate}
\usepackage[dvipdfm,bookmarks=true,colorlinks,%
           citecolor=blue,linkcolor=blue,hypertex, %
           breaklinks=true]{hyperref}

\begin{document}
\begin{CJK*}{GBK}{song}

\title{Shell-model-like approach based on cranking covariant
density functional theory: bandcrossing and shape evolution in $^{60}$Fe}

\author{Z. Shi }
\affiliation{School of Physics and Nuclear Energy Engineering,
Beihang University, Beijing 100191, China}

\author{Z. H. Zhang }
\affiliation{Mathematics and Physics Department,
North China Electric Power University, Beijing 102206, China}

\author{Q. B. Chen }
\affiliation{State Key Laboratory of Nuclear Physics and Technology,
School of Physics, Peking University, Beijing 100871, China}
\affiliation{Physik-Department, Technische Universit\"{a}t M\"{u}nchen,
D-85747 Garching, Germany}

\author{S. Q. Zhang }
\email{sqzhang@pku.edu.cn}
\affiliation{State Key Laboratory of Nuclear Physics and Technology,
School of Physics, Peking University, Beijing 100871, China}

\author{J. Meng }
\email{mengj@pku.edu.cn}
\affiliation{State Key Laboratory of Nuclear Physics and Technology,
School of Physics, Peking University, Beijing 100871, China}
\affiliation{School of Physics and Nuclear Energy Engineering,
Beihang University, Beijing 100191, China}
\affiliation{Yukawa Institute for Theoretical Physics,
Kyoto University, Kyoto 606-8502, Japan}

\begin{abstract}
The shell-model-like approach is implemented to treat the cranking many-body Hamiltonian
based on the covariant density functional theory including pairing correlations with
exact particle number conservation. The self-consistency is achieved by iterating the
single-particle occupation probabilities back to the densities and currents. As an example,
the rotational structures observed in the neutron-rich nucleus $^{60}$Fe are investigated
and analyzed. Without introducing any \emph{ad hoc} parameters, the bandheads, the
rotational spectra, and the relations between the angular momentum and rotational frequency
for the positive parity band A, and negative parity bands B and C are well reproduced.
The essential role of the pairing correlations is revealed. It is found that for band A,
the bandcrossing is due to the change of the last two occupied neutrons from the $1f_{5/2}$
signature partners to the $1g_{9/2}$ signature partners. For the two negative parity
signature partner bands B and C, the bandcrossings are due to the pseudo-crossing between
the $1f_{7/2,~5/2}$ and the $1f_{5/2,~1/2}$ orbitals. Generally speaking, the deformation
parameters $\beta$ for bands A, B, and C decrease with rotational frequency. For band A,
the deformation jumps from $\beta\sim0.19$ to $\beta\sim0.29$ around the bandcrossing. In
comparison with its signature partner band C, band B exhibits appreciable triaxial deformation.
\\

\end{abstract}

\date{\today}


\maketitle


\section{Introduction}

For the past decades, lots of novel phenomena with unexpected features in nuclear structure,
including superdeformed rotational bands~\cite{twin1986observation,ring1996relativistic},
neutron halo~\cite{tanihata1985measurements,meng1996relativistic} and giant halo~\cite{meng1998giant,
meng2006relativistic,meng2015halos}, the disappearance and occurrence of magic numbers~\cite{ozawa2000new},
magnetic and antimagnetic rotation~\cite{frauendorf1994proceedings,clark1997evidence,
frauendorf2001spontaneous,meng2013progress}, chiral doublet bands~\cite{frauendorf1997tilted,
starosta2001chiral}, and multiple chiral doublets (M$\chi$D)~\cite{meng2006possible,
ayangeakaa2013evidence,kuti2014multiple,liu2016evidence}, have attracted worldwide attentions
and challenged nuclear models aiming at a unified and microscopic interpretation of these
novel phenomena.

Starting from an effective nucleon-nucleon interaction with Lorentz invariance, the covariant
density functional theory (CDFT) naturally includes the spin-orbit coupling and has achieved
great successes in describing many nuclear phenomena in stable and exotic nuclei of the whole
nuclear chart~\cite{ring1996relativistic,meng2006relativistic,liang2015hidden,meng2015halos,
meng2016relativistic}. Based on the same functional and without introducing any additional
parameters, the CDFT can well describe the rotational excitations in nuclei by including the
cranking terms~\cite{koepf1989relativistic,madokoro2000relativistic,meng2013progress}. Up to now,
the cranking CDFT has been developed for the principal axis cranking (PAC)~\cite{koepf1991relativistic},
the tilted axis cranking (TAC)~\cite{peng2008covariant,zhao2011novel}, and also for the aplanar
TAC~\cite{madokoro2000relativistic,zhao2017multiple}. With various versions of cranking CDFT,
novel rotational phenomena including superdeformed rotational bands~\cite{koepf1991relativistic,
konig1993identical,afanasjev1996superdeformed}, magnetic~\cite{madokoro2000relativistic,
peng2008covariant,zhao2011novel,yu2012magnetic} and antimagnetic rotation~\cite{zhao2011antimagnetic,
zhao2012covariant}, linear cluster structure~\cite{zhao2015rod}, and chiral doublet
bands~\cite{zhao2017multiple}, have been investigated successfully.

Pairing correlations are essential to describe not only the nuclear ground state
properties~\cite{bohr1975nuclear,peter1980nuclear,meng2016relativistic} but also the
excited state properties~\cite{afanasjev1999cranked,niu2013beta,zhao2015impact,
meng2016relativistic,niu2017self}. Within the mean-field approximation, the pairing
correlations are usually treated by the Bardeen-Cooper-Schrieffer (BCS) approximation
or Bogoliubov transformation~\cite{peter1980nuclear}. However, the particle number is
not conserved in the standard BCS and Bogoliubov approximations. The blocking effect,
which is responsible for various odd-even differences in nuclear properties and important
for low-lying excited states, can only be approximately considered. Another difficulty
is the pairing correlation collapse with rotation~\cite{mottelson1960effect}. Moreover,
the BCS approximation can not be applied to the cranking model as the time-reversal
symmetry is broken. Although these defects can be remedied by the particle number
projection technique~\cite{dietrich1964conservation,peter1980nuclear,anguiano2001particle,
stoitsov2007variation,yao2014microscopic}, the calculation algorithm is complicated
and the simplicity is lost~\cite{peter1980nuclear}.

Shell-model-like approach (SLAP)~\cite{meng2006shell}, or originally referred as particle
number conserving (PNC) method~\cite{zeng1983particle}, treats pairing correlations and
blocking effects exactly by diagonalizing the many-body Hamiltonian in a many particle
configuration (MPC) space with conserved particle number. Based on the phenomenological
cranking Nilsson model, extensive applications for the odd-even differences in moments
of inertia~\cite{zeng1994blocking}, identical bands~\cite{liu2002microscopic,He2005the},
nuclear pairing phase transition~\cite{wu2011nuclear}, antimagnetic rotation~\cite{zhang2013nuclear,
zhang2016effects}, and high-$K$ rotational bands in the rare-earth~\cite{liu2004particle,
zhang2009particle2,zhang2009particle,li2013particle,li2016rotational,zhang2016particle,
zhang2013rotational} and actinide nuclei~\cite{he2009influence,zhang2011particle,zhang2012systematic},
have been performed. Furthermore, the SLAP has been combined with CDFT~\cite{meng2006shell,
liu2015thermodynamics}, deformed Woods-Saxon potential~\cite{molique1997fock,fu2013configuration},
and the Skyrme density functional~\cite{pillet2002pairing,liang2015configuration}. Similar
approaches to treat pairing correlations with conserved particle number can be found in
Refs.~\cite{richardson1964exact,pan1998particle,volya2001exact,jia2013particle,jia2013accuracy,
chen2014relativistic}. Based on the CDFT, the SLAP has been firstly adopted to study the
ground state properties and low-lying excited states for Ne isotopes~\cite{meng2006shell}.
In Ref.~\cite{liu2015thermodynamics}, the extension to include the temperature has been
implemented to study the heat capacity.

In this paper, the SLAP is implemented to treat the cranking many-body Hamiltonian based
on the CDFT including pairing correlations with exact particle number conservation, and
is referred as cranking CDFT-SLAP. Our aim is to investigate the rotational excitation
modes of superfluid nuclei in a fully microscopic, self-consistent, and particle number
conserved manner. As an example, the rotational spectra in the neutron-rich nucleus $^{60}$Fe
will be investigated with one of the most successful functionals PC-PK1~\cite{zhao2012covariant}.
Being a key isotope in astrophysics and cosmic nucleosynthesis, the low-lying structure
and rotational spectra in $^{60}$Fe have been investigated experimentally~\cite{warburton1977yrast,
norman1978energy,wilson2000high,deacon2007yrast} and theoretically with the projected
shell model~\cite{sun2009projected} and the large-scale shell-model~\cite{togashi2015large}.

The paper is organized as follows. In Sec.~\ref{sec1}, the theoretical frameworks for the
cranking CDFT and the SLAP  are briefly presented. The numerical details are given in
Sec.~\ref{sec2}. In Sec.~\ref{sec3}, the energy spectra and the relations between total
angular momenta and rotational frequency for the three rotational bands in $^{60}$Fe
calculated by the cranking CDFT-SLAP are presented and compared with the data. The
bandcrossing mechanisms and shape evolutions in these rotational bands are discussed.
A short summary is given in Sec.~\ref{sec4}.


\section{Theoretical framework} \label{sec1}

\subsection{Cranking covariant density functional theory}

The effective Lagrangian density for the point-coupling covariant density functional is
as follows~\cite{zhao2010new,zhao2012covariant},
\begin{align}\label{eq-lagrangian}
 \mathcal L=&\mathcal L^{\rm free}+\mathcal L^{4\rm f}+\mathcal L^{\rm hot}
             +\mathcal L^{\rm der}+\mathcal L^{\rm em} \notag\\
           =&\bar\psi(i\gamma_\mu\partial^\mu-m)\psi \notag \\
            &-\frac{1}{2}\alpha_S(\bar\psi\psi)(\bar\psi\psi)
             -\frac{1}{2}\alpha_V(\bar\psi\gamma_\mu\psi)(\bar\psi\gamma^\mu\psi)
             -\frac{1}{2}\alpha_{TV}(\bar\psi\vec{\tau}\gamma_\mu\psi)
              (\bar\psi\vec{\tau}\gamma^\mu\psi)\notag \\
            &-\frac{1}{3}\beta_S(\bar\psi\psi)^3-\frac{1}{4}\gamma_S(\bar\psi\psi)^4
             -\frac{1}{4}\gamma_V[(\bar\psi\gamma_\mu\psi)(\bar\psi\gamma^\mu\psi)]^2\notag \\
            &-\frac{1}{2}\delta_S\partial_\nu(\bar\psi\psi)\partial^\nu(\bar\psi\psi)
             -\frac{1}{2}\delta_V\partial_\nu(\bar\psi\gamma_\mu\psi)
              \partial^\nu(\bar\psi\gamma^\mu\psi)
             -\frac{1}{2}\delta_{TV}\partial_\nu(\bar\psi\vec{\tau}\gamma_\mu\psi)
              \partial^\nu(\bar\psi\vec{\tau}\gamma^\mu\psi) \notag \\
            &-\frac{1}{4}F^{\mu\nu}F_{\mu\nu}-e\frac{1-\tau_3}{2}\bar\psi\gamma^\mu\psi A_\mu,
\end{align}
which includes the free nucleon term $\mathcal L^{\rm free}$, the four-fermion point-coupling
terms $\mathcal L^{4\rm f}$, the higher-order terms $\mathcal L^{\rm hot}$ responsible for the
medium effects, the gradient terms $\mathcal L^{\rm der}$ simulating the effects of finite range,
and the electro-magnetic interaction terms $\mathcal L^{\rm em}$.

To describe the nuclear rotation, the effective Lagrangian (\ref{eq-lagrangian}) is
transformed into a rotating frame with a constant rotational frequency $\omega_x$ around
the $x$ axis~\cite{koepf1989relativistic,kaneko1993three,konig1993identical}. The equation
of motion for the nucleons derived from the rotating Lagrangian is written as
\begin{equation}\label{eq-dirac-1d}
   \hat h_0\psi_\mu = (\hat h_{\rm s.p.}+\hat h_{\rm c})\psi_\mu=\varepsilon_\mu\psi_\mu,
\end{equation}
with
\begin{align}
    \hat h_{\rm s.p.} = \bm \alpha\cdot(-i{\bm\nabla}-\bm{V})+\beta(m+S)+V^0,~~~~
    \hat h_{\rm c} = -{\omega_x\cdot\hat{j}_x},
\end{align}
where $ \hat j_x =\bm \hat l_x+\frac{1}{2} \Sigma_x$ is $x$-component of the total angular
momentum of the nucleon spinors, and $\varepsilon_\mu$ represents the single-particle Routhians
for nucleons. The relativistic fields $S({\bm r})$ and $V^\mu(\bm r)$ have the form
\begin{align}
    &S(\bm r)=\alpha_S\rho_S+\beta_S\rho^2_S+\gamma_S\rho^3_S+\delta_S\Delta\rho_S,\notag \\
    &V^0(\bm r)=\alpha_V\rho_V+\gamma_V\rho^3_V+\delta_V\Delta\rho_V+\tau_3\alpha_{TV}\rho_{TV}
                +\tau_3\delta_{TV}\Delta\rho_{TV}+e\frac{1-\tau_3}{2}{A}^0,\notag \\
    &\bm{V}({\bm r})=\alpha_V\bm{j}_V+\gamma_V(\bm{j}_V)^3+\delta_V\Delta\bm{j}_V
                    +\tau_3\alpha_{TV}\bm{j}_{TV}+\tau_3\delta_{TV}\Delta\bm{j}_{TV}
                    +e\frac{1-\tau_3}{2}{\bm A},
\end{align}
with $\rho$ and $\bm j$ respectively represent the local densities and currents,
\begin{align}\label{eq-density-curr}
    \rho_S(\bm r)&=\sum_{\mu}n_\mu\bar\psi_\mu(\bm r)\psi_\mu(\bm r), \notag \\
    \rho_V(\bm r)&=\sum_{\mu}n_\mu\psi^\dag_\mu(\bm r)\psi_\mu(\bm r),\notag \\
    \bm{j}_V(\bm r)&=\sum_{\mu}n_\mu\psi^\dag_\mu(\bm r)\bm\alpha\psi_\mu(\bm r),\notag \\
    \rho_{TV}(\bm r)&=\sum_{\mu}n_\mu\psi^\dag_\mu(\bm r)\bm\tau_3\psi_\mu(\bm r),\notag \\
    \bm{j}_{TV}(\bm r)&=\sum_{\mu}n_\mu\psi^\dag_\mu(\bm r)\bm\alpha\tau_3\psi_\mu(\bm r), \notag \\
    \rho_c(\bm r)&=\sum_{\mu}n_\mu\psi^\dag_\mu(\bm r)\frac{1-\tau_3}{2}\psi_\mu(\bm r),
\end{align}
in which $n_\mu$ is the occupation probability for each state $\mu$. The sums are taken
over the states with positive energies only, i.e., the contributions of the negative-energy
states are neglected (no-sea approximation). It is noted that the spatial components of
the electro-magnetic vector potential $\bm A$ are neglected since their contributions are
extremely small.

After solving the equation of motion (\ref{eq-dirac-1d}) self-consistently, the total energy
of the system in the laboratory is obtained as,
\begin{align}\label{etot-eq}
     E_{\rm tot}=E_{\rm kin}+E_{\rm int}+E_{\rm cou}+E_{\rm c.m.},
\end{align}
with the energies of kinetic part,
\begin{align}
     E_{\rm kin}&=\int d^3{\bm r}\sum_{\mu}n_\mu\psi^\dag_\mu[\bm{\alpha}\cdot\bm p+\beta m]\psi_\mu,
\end{align}
the interaction part,
\begin{align}
     E_{\rm int}&=\int d^3{\bm r}\left\{\frac{1}{2}\alpha_S\rho^2_S+\frac{1}{3}\beta_S\rho^3_S
                +\frac{1}{4}\gamma_S\rho^4_S+\frac{1}{2}\delta_S\rho_S\Delta\rho_S\right. \notag\\
                &~~~~~~+\frac{1}{2}\alpha_V(\rho^2_V-{\bm j}\cdot{\bm j})
                +\frac{1}{2}\alpha_{TV}(\rho^2_{TV}-{\bm j}_{TV}\cdot{\bm j}_{TV}) \notag\\
                &~~~~~~+\frac{1}{4}\gamma_V(\rho^2_V-{\bm j}\cdot{\bm j})^2
                +\frac{1}{2}\delta_V(\rho_V\Delta\rho_V-{\bm j}\Delta{\bm j}) \notag\\
                &~~~~~~+\left.\frac{1}{2}\delta_{TV}(\rho_{TV}\Delta\rho_{TV}
                -{\bm j}_{TV}\Delta{\bm j}_{TV})\right\},
\end{align}
the electro-magnetic part,
\begin{align}
    E_{\rm cou}=\int d^3{\bm r}\frac{1}{2}eA_0\rho_c,
\end{align}
and the center-of-mass (c.m.) correction part,
\begin{align}
    E_{\rm c.m.}=-\frac{\langle {\hat{\bm P}}^2_{\rm c.m.}\rangle}{2mA},
\end{align}
with the mass number $A$ and the total momentum in the center-of-mass frame
$ {\hat{\bm P}}_{\rm c.m.}=\sum_i {\hat{\bm p}}_i$.

The Dirac equation (\ref{eq-dirac-1d}) can be solved by expanding the nucleon spinors in a
complete set of basis states. In the present work, a three-dimensional harmonic oscillator
(3DHO) basis in Cartesian coordinates~\cite{koepf1988has,dobaczewski1997solution,yao2006time,
peng2008covariant,nikvsic2009beyond} with good signature quantum number is adopted,
\begin{align} \label{3dho-nega}
    \Phi_{\xi+}({\bm r}, {\bm s})
    =\langle {\bm r}, {\bm s}|\xi\alpha=+\rangle
   &=\phi_{n_x}\phi_{n_y}\phi_{n_z}\frac{i^{n_y}}{\sqrt{2}}(-1)^{n_z+1}
    \left(
    \begin{array}{ccc}
    1\\
    (-1)^{n_y+n_z}
    \end{array}
    \right), \\
    \Phi_{\xi-}({\bm r}, {\bm s})
    =\langle {\bm r}, {\bm s}|\xi\alpha=-\rangle
   &=\phi_{n_x}\phi_{n_y}\phi_{n_z}\frac{i^{n_y}}{\sqrt{2}}
    \left(
    \begin{array}{ccc}
    1\\
    (-1)^{n_y+n_z+1}
    \end{array}
    \right),
    \label{3dho-posi}
\end{align}
which correspond to the eigenfunctions of the signature operation with the positive $(\alpha=+1/2)$
and negative $(\alpha=-1/2)$ eigenvalues, respectively. The $n_x$, $n_y$ and $n_z$ are
the harmonic oscillator quantum numbers in $x$, $y$, and $z$ directions and $\phi_{n_x}$,
$\phi_{n_y}$ and $\phi_{n_z}$ are the corresponding eigenstates. The phase factor $i^{n_y}$
is added in order to get real matrix elements for the Dirac equation~\cite{meng2013progress}.
Furthermore, under the time-reversal operation $\mathcal {\hat T}= -i\sigma_y \hat K$, this
3DHO basis fulfills the following properties,
\begin{align}\label{3dho-feature}
    \mathcal {\hat T}\Phi_{\xi+}({\bm r}, {\bm s})=\Phi_{\xi-}({\bm r}, {\bm s}),~~
    \mathcal {\hat T}\Phi_{\xi-}({\bm r}, {\bm s})=-\Phi_{\xi+}({\bm r}, {\bm s}).
\end{align}
It means that under a proper phase factor, the $\Phi_{\xi+}$ and $\Phi_{\xi-}$ are a pair of
time-reversal states with the same quantum numbers $n_x$, $n_y$ and $n_z$.

\subsection{Shell-model-like approach}

The cranking many-body Hamiltonian with pairing correlations reads,
\begin{align}\label{hami-eq-crank}
   \hat H=\hat H_0+\hat H_{\rm pair}.
\end{align}
The one-body Hamiltonian $\hat H_0=\sum\hat h_0$ with $\hat h_0$ given in Eq.~(\ref{eq-dirac-1d}).
The monopole pairing Hamiltonian $\hat H_{\rm pair}$ is used
\begin{align}\label{eq-hpair-common}
    \hat H_{\rm pair}=-G\sum_{\xi,\eta>0}^{\xi\neq\eta}\hat\beta^\dag_{\xi}\hat\beta^\dag_{\bar\xi}
                        \hat\beta_{\bar\eta}\hat\beta_{\eta},
\end{align}
where $G$ is the effective pairing strength, $\bar\xi$ ($\bar\eta$) labels the time-reversal
state of $\xi$ ($\eta$), and $\xi\neq\eta$ means that the self-scattering for the nucleon pairs
is forbidden~\cite{meng2006shell}.

The one-body Hamiltonian $\hat H_0$ in the 3DHO basis (\ref{3dho-nega})-(\ref{3dho-posi})
can be written as
\begin{align}\label{eq-h0-base}
    \hat H_0 = \sum_{\xi\eta,\alpha}h_{\xi\alpha,\eta\alpha}
               \hat\beta^\dag_{\xi\alpha}\hat\beta_{\eta\alpha}.
\end{align}
Here $h_{\xi\alpha,\eta\alpha}$ is the matrix element of $\hat h_0$ between states $|\xi\alpha\rangle$
and $|\eta\alpha\rangle$. Accordingly, the pairing Hamiltonian $\hat H_{\rm pair}$ in the 3DHO basis
can be written as
\begin{align}\label{eq-hpair-base}
    \hat H_{\rm pair} =-G\sum_{\xi,\eta>0}^{\xi\neq\eta}\hat\beta^\dag_{\xi+}\hat\beta^\dag_{\xi-}
                        \hat\beta_{\eta-}\hat\beta_{\eta+}.
\end{align}

The idea of SLAP is to diagonalize the many-body Hamiltonian in a properly truncated MPC
space with exact particle number~\cite{zeng1983particle}. One can diagonalize the cranking
many-body Hamiltonian (\ref{hami-eq-crank}) in the MPC space constructed from the
single-particle states either in the CDFT or in the cranking CDFT. The latter is expected
to achieve the same accuracy with smaller MPC space.

Diagonalizing the one-body Hamiltonian $\hat H_0$ (\ref{eq-h0-base}) in the basis
$|\xi\alpha\rangle$ (\ref{3dho-nega})-(\ref{3dho-posi}), one can obtain the single-particle
Routhian $\varepsilon_{\mu\alpha}$ and the corresponding eigenstate $|\mu\alpha\rangle$ for
each level $\mu$ with the signature $\alpha$, namely,
\begin{align}
    \hat H_0 =\sum_{\mu\alpha} \varepsilon_{\mu\alpha}\hat b^\dag_{\mu\alpha}\hat b_{\mu\alpha},~~~~
    |\mu\alpha\rangle = \sum_\xi C_{\mu\xi}(\alpha)|\xi\alpha\rangle.
\end{align}
From the real expansion coefficient $C_{\mu\xi}(\alpha)$, the transformation between the
operators $\hat b^\dag_{\mu\alpha}$ and $\hat\beta^\dag_{\xi\alpha}$ can be expressed as,
\begin{align}\label{eq-state-base}
    \hat b^\dag_{\mu\alpha}=\sum_\xi C_{\mu\xi}(\alpha)\hat\beta^\dag_{\xi\alpha}, ~~~~
    \hat\beta^\dag_{\xi\alpha}=\sum_\mu C_{\mu\xi}(\alpha)\hat b^\dag_{\mu\alpha}.
\end{align}

In the $|\mu\alpha\rangle$ basis, the pairing Hamiltonian $\hat H_{\rm pair}$ can be written as,
\begin{align}\label{eq-hpair}
    \hat H_{\rm pair} = -G \sum_{\mu\mu^\prime\nu\nu^\prime}\sum_{\xi,\eta>0}^{\xi\neq\eta}
                        C_{\mu\xi}(+)C_{\mu^\prime\xi}(-)C_{\nu\eta}(-)C_{\nu^\prime\eta}(+)
                        \hat b^\dag_{\mu+}\hat b^\dag_{\mu^\prime-}\hat b_{\nu-}\hat b_{\nu^\prime+}.
\end{align}

From the single-particle Routhian $\varepsilon_{\mu\alpha}$ and the corresponding eigenstate
$|\mu\alpha\rangle$ (briefly denoted by $|\mu\rangle$), the MPC $|i\rangle$ for an $n$-particle
system can be constructed as~\cite{zeng1994reduction}
\begin{align}
   |i\rangle= |\mu_1\mu_2\cdots\mu_n\rangle
            = \hat b^\dag_{\mu_1}\hat b^\dag_{\mu_2}\cdots \hat b^\dag_{\mu_n}|0\rangle.
\end{align}
The parity $\pi$, signature $\alpha$, and the corresponding configuration energy for each
MPC are obtained from the occupied single-particle states.

The eigenstates for the cranking many-body Hamiltonian are obtained by diagonalization in
the MPC space,
\begin{align}
    |\Psi\rangle = \sum_i C_i |i\rangle,
\end{align}
with $C_i$ the expanding coefficients.

The occupation probability $n_\mu$ for state $\mu$ is,
\begin{align}\label{eq-occupy}
    n_\mu=\sum_i |C_i|^2P_{i\mu},~~
    P_{i\mu}=\left\{
\begin{array}{cl}
  1, &~~~~|i\rangle~{\rm contains~|\mu\rangle},\\
  0, &~~~~{\rm otherwise}.
\end{array}
\right.
\end{align}
The occupation probabilities will be iterated back into the densities and currents in
Eq.~(\ref{eq-density-curr}) to achieve self-consistency~\cite{meng2006shell}.

It is noted that, for the total energy in CDFT (\ref{etot-eq}), the pairing energy due to
the pairing correlations should be taken into account,
$E_{\rm pair} = \langle\Psi|\hat H_{\rm pair}|\Psi\rangle$.


\section{Numerical details}\label{sec2}

\begin{figure*}[t!]
  \centering
  \includegraphics[scale=0.45,angle=0]{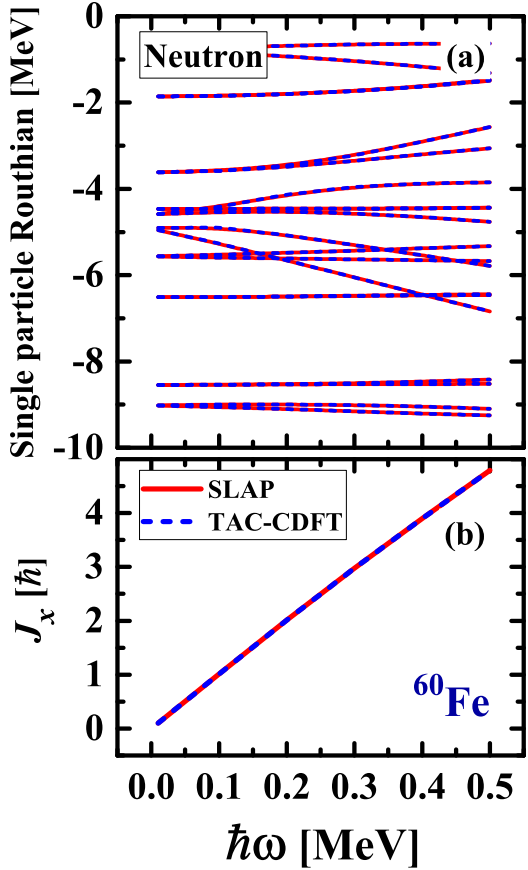}
  \caption{(Color online) The neutron single-particle Routhian (a) and the alignment along
  the   rotational axis $J_x$ (b) as functions of the rotational frequency in $^{60}$Fe
  calculated by   the cranking CDFT-SLAP with PC-PK1~\cite{zhao2010new}, in comparison with
  the TAC-CDFT~\cite{zhao2011novel} calculations with tilted angle $\theta=0^\circ$. }
\label{fe60-check-self}
\end{figure*}

As mentioned before, the cranking many-body Hamiltonian (\ref{hami-eq-crank}) can be
diagonalized in the MPC space constructed from the single-particle states either in the
CDFT or in the cranking CDFT. The latter is expected to achieve the same accuracy with
smaller MPC space.

In the following, the validity for diagonalizing the cranking many-body Hamiltonian
(\ref{hami-eq-crank}) in MPC space constructed from the single-particle states in the
cranking CDFT, namely cranking CDFT-SLAP, will be checked.

In the present cranking CDFT-SLAP calculations for $^{60}$Fe, the point-coupling density
functional PC-PK1~\cite{zhao2010new} is used in the particle-hole channel and the monopole
pairing interaction is adopted in the particle-particle channel. The equation of motion
(\ref{eq-dirac-1d}) is solved by expanding the Dirac spinor in terms of the three dimensional
harmonic oscillator basis (\ref{3dho-nega})-(\ref{3dho-posi}) with 10 major shells. For
both neutron and proton, the dimensions of the MPC space are chosen as 800, which correspond
to the energy cutoffs $E_{\rm c}\approx12.1$ and $\approx18.5$ MeV, respectively. The effective
pairing strengths are 0.8 MeV for both neutron and proton by reproducing the experimental
odd-even mass differences. Increasing the number of major shells from 10 to 12, the change
of the total energy is within 0.1\%. Increasing the dimension of the MPC space from 800
to 1200 and adjusting the effective pairing strength accordingly, the change of the total
energy is within 0.1\%. In the present calculation, there is no free parameter.

The validity for cranking CDFT-SLAP at the rotational frequency $\hbar\omega=0.0$ MeV is
confirmed by reproducing the results in Ref.~\cite{meng2006shell}, indicating that the
pairing correlations have been taken into account correctly.

The validity for cranking CDFT-SLAP is also checked against the TAC-CDFT~\cite{zhao2011novel}
calculation with the pairing correlations switching off. The neutron single-particle Routhian
and the alignment along the rotational axis $J_x=\langle\Psi|\hat J_x|\Psi\rangle$ as functions
of the rotational frequency in $^{60}$Fe calculated by the cranking CDFT-SLAP are shown in
Fig.~\ref{fe60-check-self}, in comparison with the TAC-CDFT~\cite{zhao2011novel} calculations
with tilted angle $\theta=0^\circ$. Satisfactory agreement is found with the differences
less than $10^{-4}$ MeV for the neutron single-particle Routhian and $10^{-4}\hbar$ for $J_x$.


\section{Results and Discussion}\label{sec3}

Three rotational bands of the neutron-rich nucleus $^{60}$Fe have been observed in
Ref.~\cite{deacon2007yrast}, including the yrast band with positive parity (labeled as band A)
and two negative-parity signature partner bands with similar intensity starting from $6^-$
and $5^-$ states (labeled as bands B and C), respectively. As both parity and signature are
good quantum numbers, the cranking many-body Hamiltonian (\ref{hami-eq-crank}) can be
diagonalized in the corresponding MPC space. The yrast bands thus obtained for different
parity and signature are compared with the observed bands A, B and C.

\subsection{Energy spectra and $I-\omega$ relations}

\begin{figure*}[h!]
  \centering
  \includegraphics[scale=0.6,angle=0]{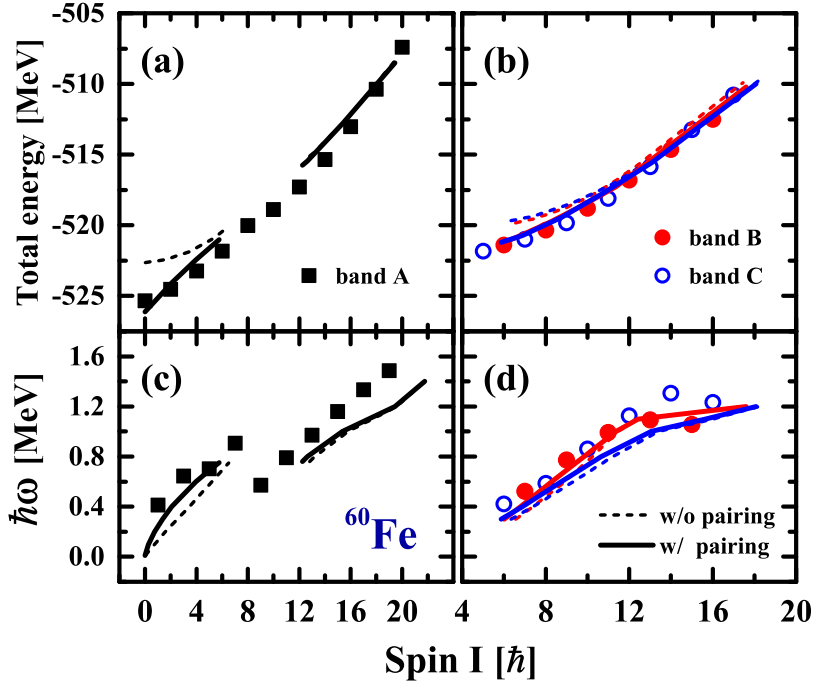}
  \caption{(Color online) The total energies (a)-(b) and the rotational frequencies (c)-(d)
  as functions of the spin for the positive parity band A, and negative parity signature
  partner bands B and C in $^{60}$Fe   calculated by the cranking CDFT-SLAP with and
  without pairing, in comparison with the data available~\cite{deacon2007yrast}.}
\label{fe60-i-omega-energy}
\end{figure*}

At a given rotational frequency, the eigenstate $|\Psi\rangle$ of the cranking many-body
Hamiltonian (\ref{hami-eq-crank}) can be obtained by diagonalization in the MPC space. By
adding the pairing energy $E_{\rm pair} = \langle\Psi|\hat H_{\rm pair}|\Psi\rangle$ to
Eq.~(\ref{etot-eq}), the total energy of the system can be obtained. The corresponding
spin $I$ can be obtained through $J_x=\langle\Psi|\hat J_x|\Psi\rangle=\sqrt{I(I+1)}$.

In Fig.~\ref{fe60-i-omega-energy}, the total energies and the rotational frequencies are
shown as functions of the spin for the positive parity band A, and negative parity signature
partner bands B and C in $^{60}$Fe calculated by the cranking CDFT-SLAP with and without
pairing correlations, in comparison with the available data.

In Figs.~\ref{fe60-i-omega-energy}(a)-\ref{fe60-i-omega-energy}(b), the cranking CDFT-SLAP
calculations well reproduce the energy spectra for bands A, B and C without introducing
any \emph{ad hoc} parameters. Switching off the pairing correlations, the deviations appear
for the low-spin regions, in particular for band A.

In Figs.~\ref{fe60-i-omega-energy}(c)-\ref{fe60-i-omega-energy}(d), the cranking CDFT-SLAP
calculations well reproduce the $I-\omega$ relations including the bandcrossings for bands
A, B and C. Switching off the pairing correlations, the deviations appear for the low-spin
region for band A.

\begin{figure*}[h!]
  \centering
  \includegraphics[scale=0.5,angle=0]{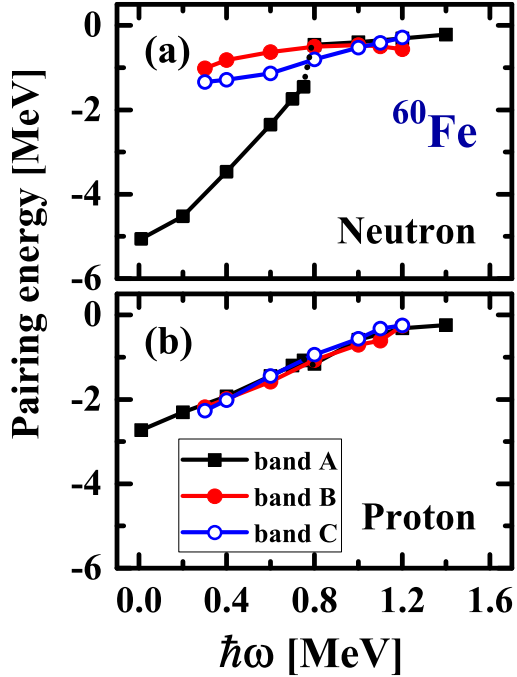}
    \caption{(Color online) The pairing energies as functions of the rotational frequency
    for neutron (a) and proton (b) in the positive parity band A, and negative parity
    signature partner bands B and C.}
\label{fe60-epair}
\end{figure*}

In Fig.~\ref{fe60-epair}, the pairing energies as functions of the rotational frequency for
neutron and proton are shown for the positive parity band A, and negative parity signature
partner bands B and C. Generally, the pairing energies decrease with rotational frequency,
but without pairing correlation collapse. This is one of advantages of the SLAP.

For neutron, as seen in Fig.~\ref{fe60-epair}(a), the pairing energy in band A changes rapidly
from $\sim -5.0$ MeV near the bandhead to $\sim -1.5$ MeV at $\hbar\omega\approx0.75$ MeV,
where the bandcrossing occurs. After bandcrossing, it changes similarly as bands B and C.
In comparison with band A, the pairing energies in bands B and C are relatively small because
the neutron pair in the $1f_{5/2,~3/2}$ orbitals is broken (see in the following).

For proton, as seen in Fig.~\ref{fe60-epair}(b), the pairing energies change smoothly and
similarly as functions of rotational frequency for bands A, B and C, which suggest that
the proton configurations are the same. In comparison with the neutron, the suppressed
pairing correlations for  proton are due to the lower level density (see in the following).

The excellent agreements with the observed energy spectra and $I-\omega$ relations indicate
that the cranking CDFT-SLAP correctly treats the pairing correlations and mean-field involved.
From the calculations, one can pin down the corresponding configurations and examine the
mechanism for bandcrossing.

\subsection{Single-particle Routhians}

\begin{figure*}[h!]
  \centering
  \includegraphics[scale=0.65,angle=0]{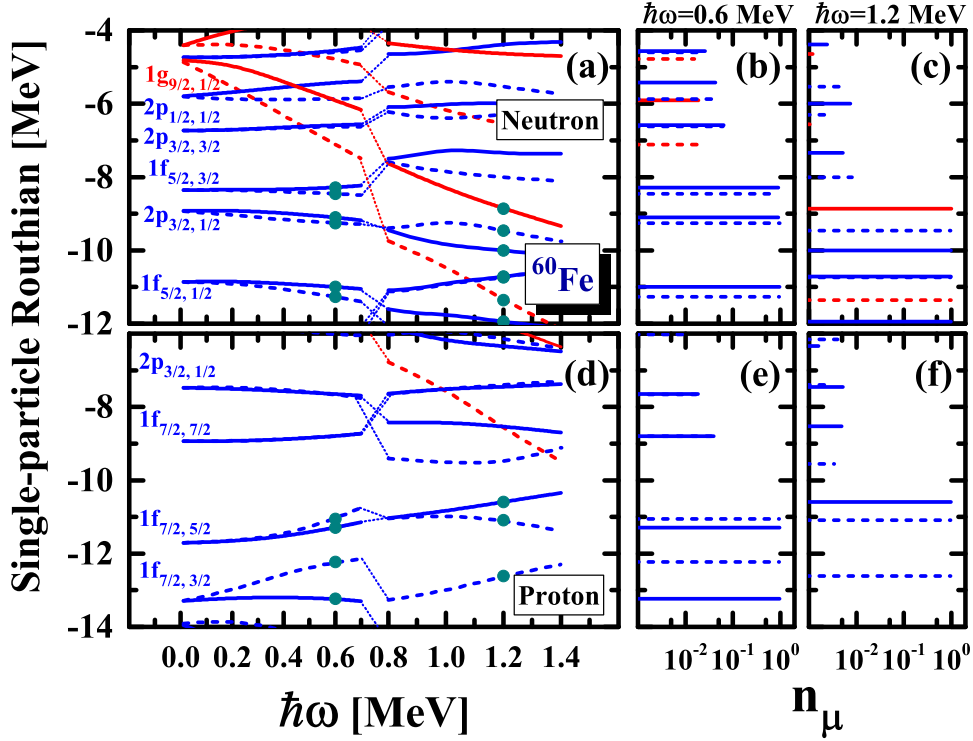}
  \caption{(Color online) The neutron (a) and proton (d) single-particle Routhians as functions
  of the rotational frequency for band A in $^{60}$Fe. Each orbital is labeled by the corresponding
  spherical quantum number of its main component. The positive (negative) parity levels are
  denoted by red (blue) lines. The signature $\alpha=+1/2$ ($-1/2$) levels are denoted by
  solid (dash) lines. The solid circles denote the occupied orbitals, and the corresponding
  occupation probabilities $n_\mu$ are given in the right two columns.}
\label{fe60-cspl-banda}
\end{figure*}

\begin{figure*}[h!]
  \centering
  \includegraphics[scale=0.65,angle=0]{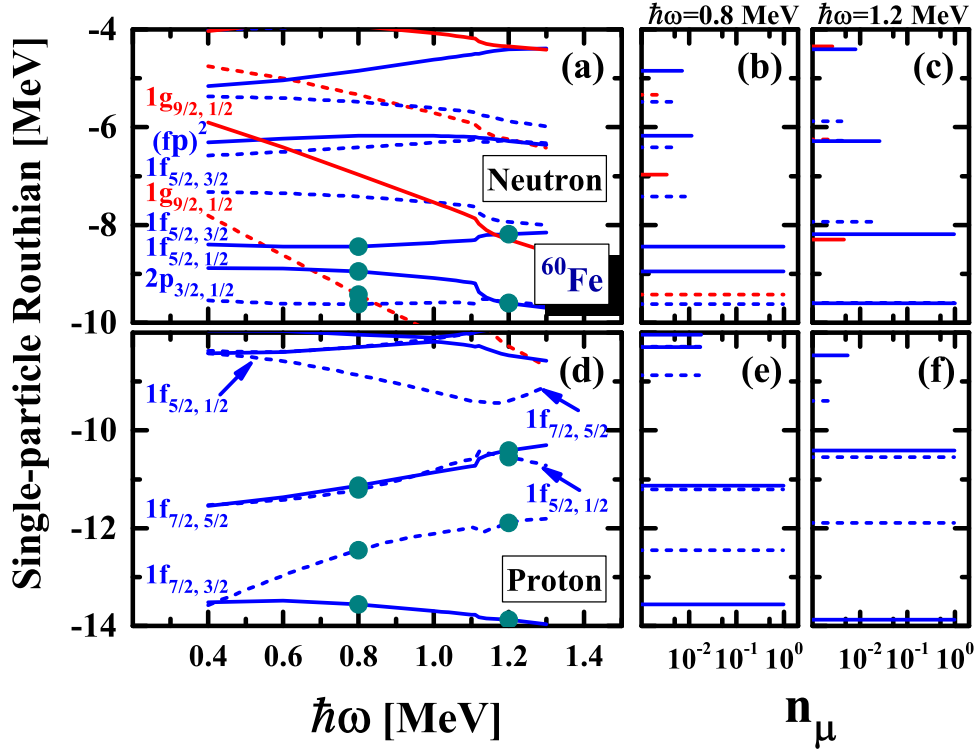}
    \caption{(Color online) Same as Fig.~\ref{fe60-cspl-banda} but for band B.}
\label{fe60-cspl-b}
\end{figure*}

To explore the mechanism of the observed bandcrossings, in Figs.~\ref{fe60-cspl-banda}
and~\ref{fe60-cspl-b}, the single-particle Routhians as functions of the rotational frequency
for bands A and B are shown.

As shown in Fig.~\ref{fe60-cspl-banda}(a), the occupied neutron orbitals for band A are
changed around $\hbar\omega\approx0.75$ MeV. A discontinuity is observed in the neutron
single-particle Routhians. Using the single-particle level tracking technique~\cite{meng2006possible},
the levels with the largest overlap ($>0.9$) before and after the discontinuity are connected.
It is found that the last two occupied neutrons change from the $1f_{5/2}$ signature partners
to the $1g_{9/2}$ signature partners. The occupation probabilities of the two $1g_{9/2}$
signature partners change from less than $10^{-1}$ at $\hbar\omega=0.6$ MeV
(c.f. Fig.~\ref{fe60-cspl-banda}(b)) to nearly 1 at $\hbar\omega=1.2$ MeV
(c.f. Fig~\ref{fe60-cspl-banda}(c)). This configuration change for band A results from the
rapid decrease of the neutron $1g_{9/2}$ orbitals with rotational frequency. For proton in
band A, as shown in Figs.~\ref{fe60-cspl-banda}(d)-\ref{fe60-cspl-banda}(f), the occupation
probabilities change smoothly. The discontinuity in the proton single-particle Routhians
results from the change of the mean-field due to the neutron bandcrossing around
$\hbar\omega\approx0.75$ MeV. Hence, the configuration for band A after bandcrossing can be
assigned as $\nu (1g_{9/2})^2(1f_{5/2})^{-2}$, which is in consistent with the assignment
by the shell model~\cite{deacon2007yrast} and the projected shell model~\cite{sun2009projected}.

For neutron in band B, as shown in Figs.~\ref{fe60-cspl-b}(a)-\ref{fe60-cspl-b}(c), the
occupation probabilities change smoothly, and the neutron configuration can be assigned
as $\nu (g_{9/2})^1(1f_{5/2})^{-1}$. For proton in band B, as shown in Figs.~\ref{fe60-cspl-b}(d)
-\ref{fe60-cspl-b}(f), a pseudo-crossing is seen between the $1f_{7/2,~5/2}$ and the
$1f_{5/2,~1/2}$ orbitals at $\hbar\omega\approx1.1$ MeV. The occupation probability of the
proton $1f_{5/2,~1/2}$ orbital change from about $10^{-2}$ to nearly $1$, while that of
the $1f_{7/2,~5/2}$ changes from nearly 1 to less than $10^{-2}$.

As band C is the signature partner of band B, its detailed discussions of the neutron and
proton single-particle Routhians and occupation probabilities are not shown here.

\subsection{Angular momentum components}

\begin{figure*}[t!]
  \centering
  \includegraphics[scale=0.95,angle=0]{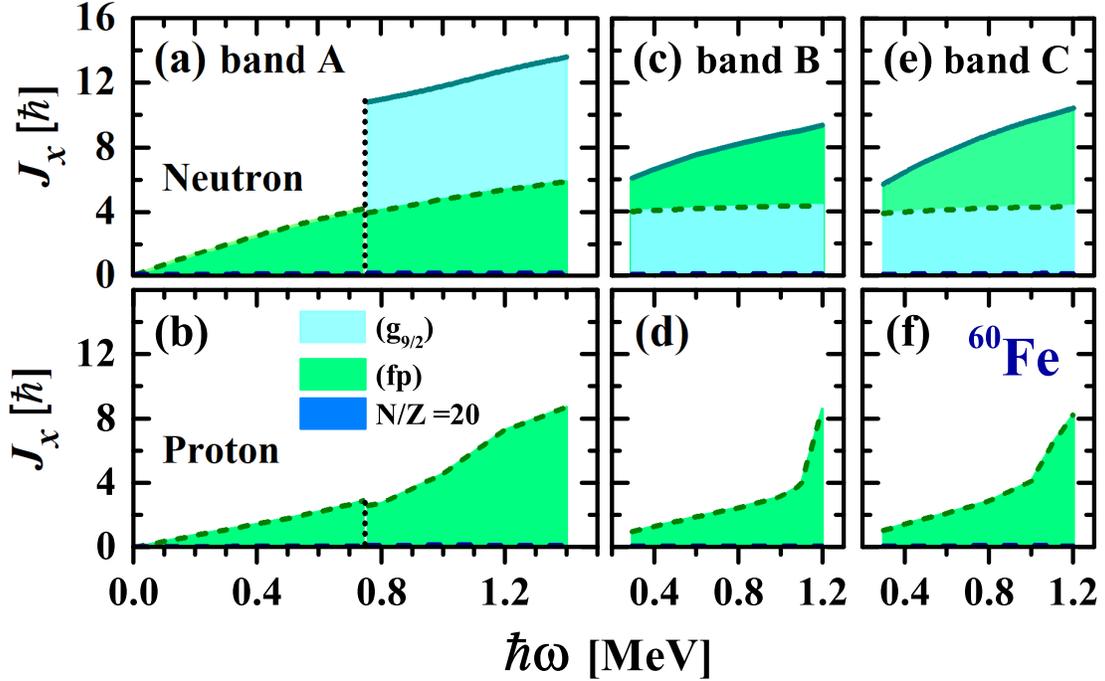}
    \caption{(Color online) The contributions from the neutron and proton $1g_{9/2}$, $(fp)$
    and $N/Z=20$ shells to the angular momentum $J_x$ as functions of the rotational frequency
    for the positive parity band A (a)-(b), and negative parity signature partner bands B (c)-(d)
    and C (e)-(f).}
\label{fe60-angular-all}
\end{figure*}

In the present fully self-consistent and microscopic cranking CDFT-SLAP calculation, the
angular momentum can be calculated from the single neutron and proton orbitals. In
Fig.~\ref{fe60-angular-all}, the contributions from the neutron and proton $1g_{9/2}$, $(fp)$
and $N/Z=20$ shells to the angular momentum $J_x$ for bands A, B and C are shown.

For all bands, both the $N=20$ and $Z=20$ shells do not contribute (core $^{40}$Ca is inert),
and only nucleons in the $(fp)$ shells and $1g_{9/2}$ orbitals contribute.

For neutron in band A, as shown in Fig.~\ref{fe60-angular-all}(a), the contributions from
the $(fp)$ shells change smoothly. After bandcrossing at $\hbar\omega\approx0.75$ MeV, the
contributions from the $1g_{9/2}$ orbitals are switched on, which produce a dramatic change
around $6\hbar$. For proton, as shown in Fig.~\ref{fe60-angular-all}(b), the contributions
are mainly from the $(fp)$ shells, which have a kink around bandcrossing but change smoothly
before and after.

For neutron in band B, as shown in Fig.~\ref{fe60-angular-all}(c), the contributions from
the $(fp)$ shells change smoothly. In contrast, the contribution from the $1g_{9/2}$ orbital
keeps nearly unchanged ($\sim4\hbar$) due to its high-$j$ low-$\Omega$ character. For proton,
as shown in Fig.~\ref{fe60-angular-all}(d), the contributions from the $(fp)$ shells increase
smoothly with the rotational frequency but much faster after $\hbar\omega=1.1$ MeV. This faster
increase is due to the pseudo-crossing between the orbitals $1f_{7/2,~5/2}$ and $1f_{5/2,~1/2}$,
as shown in Fig.~\ref{fe60-cspl-b}(b).

For neutron and proton in band C, as shown in Figs.~\ref{fe60-angular-all}(e) and
\ref{fe60-angular-all}(f), the contributions of the angular momenta can be explained similarly
as its signature partner band B.

\subsection{Shape evolution with rotation}

\begin{figure*}[t!]
  \centering
  \includegraphics[scale=0.45,angle=0]{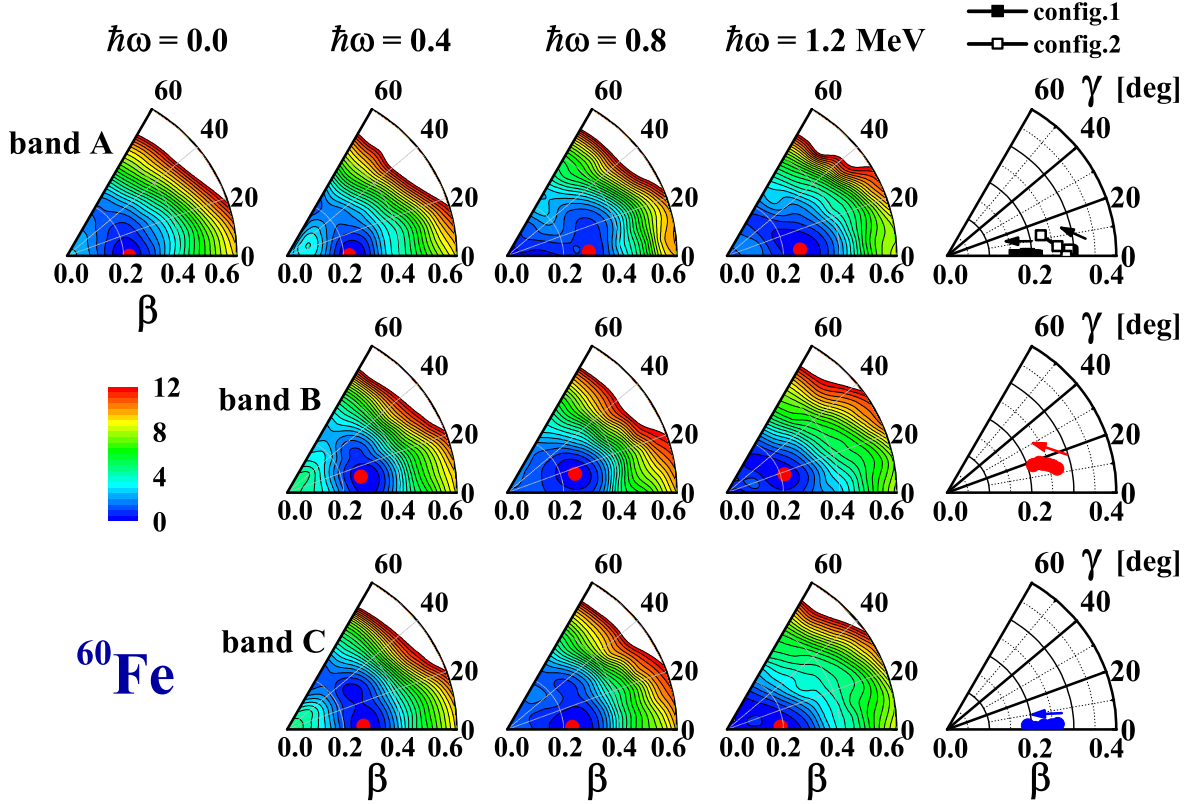}
    \caption{(Color online) The total Routhian surfaces for the positive parity band A
    (upper panels) at $\hbar\omega=0.0,~0.4,~0.8$ and 1.2 MeV, and negative parity bands
    B (middle panels) and C (lower panels) at $\hbar\omega=0.4,~0.8$ and 1.2 MeV. The red
    dot in the energy surface denotes the minimum. The energy difference between the
    neighboring contour lines is 0.5 MeV. The evolution of the deformation parameters $\beta$
    and $\gamma$ with the rotational frequency are shown in the right column. For band A, the
    configurations before and after the bandcrossing are denoted as config.1 and config.2,
    respectively.}
\label{fe60-trs-deformation}
\end{figure*}

To investigate the shape evolution with rotation in $^{60}$Fe, the total Routhian surfaces
(TRSs) for the positive parity band A (upper panels) at $\hbar\omega=0.0,~0.4,~0.8$ and 1.2 MeV,
and negative parity signature partner bands B (middle panels) and C (lower panels) at
$\hbar\omega=0.4,~0.8$ and 1.2 MeV are shown in Fig.~\ref{fe60-trs-deformation}. The evolution
of the deformation parameters $\beta$ and $\gamma$ with the rotational frequency are shown
in the right column. For band A, the configurations before and after the bandcrossing are
denoted as config.1 and config.2, respectively.

For band A, as mentioned before, the bandcrossing occurs at $\hbar\omega\approx0.75$ MeV.
Before the bandcrossing, the deformation parameters $(\beta,~\gamma)$ of the TRS minimum
at $\hbar\omega=0.0$ MeV are around $(0.21,~0^\circ)$. With the increase of the rotational
frequency, the deformation parameter $\beta$ decreases but the potential becomes more
rigid. After the bandcrossing, the deformation parameters of the TRS minimum at
$\hbar\omega=0.8$ MeV are around $(0.29,~0^\circ)$. The dramatic change of the $\beta$
results from the deformation driving effect of the neutron $1g_{9/2}$ orbital. With the
increase of the rotational frequency, the deformation parameter $\beta$ decreases but
$\gamma$ increases. The potential becomes more rigid with $\beta$ but softer with $\gamma$.
The deformation parameters of the TRS minimum at $\hbar\omega=1.4$ MeV are around
$(0.23,~12^\circ)$.

For band B, the deformation parameters of the TRS minimum at $\hbar\omega=0.4$ MeV are
around $(0.27,~12^\circ)$. With the increase of the rotational frequency, the deformation
parameter $\beta$ decreases but $\gamma$ increases, and the potential becomes softer.
The deformation parameters of the TRS minimum at $\hbar\omega=1.2$ MeV are around
$(0.21,~18^\circ)$.

Although bands B and C are signature partner bands, there is no triaxial deformation
in band C. With the increase of the rotational frequency, the deformation of the TRS
minimum decreases from $\beta=0.27$ at $\hbar\omega=0.4$ MeV to $\beta=0.19$ at
$\hbar\omega=1.2$ MeV.


\section{Summary}\label{sec4}

In summary, the shell-model-like approach is implemented to treat the cranking many-body
Hamiltonian based on the covariant density functional theory including pairing correlations
with exact particle number conservation, referred as cranking CDFT-SLAP. The self-consistency
is achieved by iterating the single-particle occupation probabilities back to the densities
and currents.

As an example, the rotational spectra observed in the neutron-rich nucleus $^{60}$Fe, including
the positive parity band A and two negative parity signature partner bands B and C, are
investigated and analyzed. Without introducing any \emph{ad hoc} parameters, the bandheads,
the rotational spectra, and the relations between the angular momentum and rotational frequency
for bands A, B, and C are well reproduced. It is found that pairing correlations are important
to describe these quantities, especially for the low-spin part. By examining the single-particle
Routhians, the occupation probabilities and the contributions from the $1g_{9/2}$, $(fp)$ and
$N/Z=20$ shells to the angular momentum, the mechanisms of the bandcrossings are analyzed and
discussed in detail. It is found that for band A, the bandcrossing is due to the change of the
last two occupied neutrons from the $1f_{5/2}$ signature partners to the $1g_{9/2}$ signature
partners. For the two negative parity signature partner bands B and C, the bandcrossings are
due to the pseudo-crossing between the $1f_{7/2,~5/2}$ and the $1f_{5/2,~1/2}$ orbitals. The
shape evolutions with rotation are investigated from the total Routhian surfaces. For band A,
the deformation parameter $\beta$ decreases with rotational frequency before and after the
bandcrossing. A dramatic change of $\beta$ is observed around the bandcrossing at the frequency
$\hbar\omega\approx0.75$ MeV, which results from the deformation driving effect of the neutron
$1g_{9/2}$ orbital. For band B, the deformation evolves from $(0.27,~12^\circ)$ at
$\hbar\omega=0.4$ MeV to $(0.21,~18^\circ)$ at $\hbar\omega=1.2$ MeV. For band C, there is no
triaxial deformation, and the deformation evolves from $\beta=0.27$ at $\hbar\omega=0.4$ MeV
to $\beta=0.19$ at $\hbar\omega=1.2$ MeV.

\begin{acknowledgments}
ZS is indebted to Fangqi Chen, Jing Peng, Zhengxue Ren, Yakun Wang, Yuanyuan Wang, Xinhui Wu,
Binwu Xiong and Pengwei Zhao for the fruitful discussions. This work was partly supported by the
Chinese Major State 973 Program (Grant No. 2013CB834400), the National Natural Science Foundation
of China (NSFC) under Grants No. 11335002, No. 11505058, No. 11621131001, No. 11775026, No. 11375015,
No. 11461141002, and the Deutsche Forschungsgemeinschaft (DFG) and NSFC through funds provided to
the Sino-German CRC 110 ``Symmetries and the Emergence of Structure in QCD''.
\end{acknowledgments}


\end{CJK*}
\end{document}